\documentclass{IEEEtran}

\usepackage{amsmath}
\usepackage{cite}
\usepackage{color}
\usepackage{mdwtab}
\usepackage{subfigure}
\usepackage{placeins}
\usepackage{psfrag}
\usepackage{graphicx}
\usepackage{epsfig}
\usepackage{epstopdf}
\usepackage[latin1]{inputenc}
\usepackage{amssymb}
\usepackage{amsthm}
\usepackage{comment}
\usepackage{setspace}

\theoremstyle{plain}
\newtheorem{theorem}{Theorem}

\theoremstyle{remark}
\newtheorem{lemma}{Lemma}

\IEEEoverridecommandlockouts 

\begin{document}

\title{Fundamental Limits on Throughput Capacity in Information-Centric Networks}

\author{Bita~Azimdoost,
        Cedric~Westphal,~\IEEEmembership{Senior Member,~IEEE,}\\
		and Hamid~R.~Sadjadpour,~\IEEEmembership{Senior Member,~IEEE,} 
   
\thanks{B. Azimdoost and H. R. Sadjadpour are with the Department of Electrical Engineering,
University of California, Santa Cruz, 1156 High Street, Santa Cruz,
CA 95064, USA (e-mail:\{bazimdoost, hamid\}@soe.ucsc.edu)}
\thanks{Bita Azimdoost was with Huawei Innovation Center, Santa Clara, CA 95050, USA, as an intern while working on this paper.}
\thanks{Cedric Westphal is with the Department of Computer Engineering,
University of California, Santa Cruz, 1156 High Street, Santa Cruz,
CA 95064, USA and Huawei Innovation Center, Santa Clara, CA 95050, USA (e-mail:cedric.westphal@huawei.com)}}     


\maketitle
\vspace{-0.4in}
\begin{abstract}
Wireless information-centric networks consider storage as one of the network primitives, and propose to cache data within the network in order to improve latency and reduce bandwidth consumption. We study the throughput capacity and latency in an information-centric network when the data cached in each node has a limited lifetime. The results show that with some fixed request and cache expiration rates, the order of the data access time does not change with network growth, and the maximum throughput order is not changing with the network growth in grid networks, and is inversely proportional to the number of nodes in one cell in random networks. Comparing these values with the corresponding throughput and latency with no cache capability (throughput inversely proportional to the network size, and latency of order $\sqrt{n}$ and the inverse of the transmission range in grid and random networks, respectively), we can actually quantify the asymptotic advantage of caching. Moreover, we compare these scaling laws for different content discovery mechanisms and illustrate that not much gain is lost when a simple path search is used. 
\end{abstract}

\IEEEpeerreviewmaketitle
\vspace{-0.6em}
\section{Introduction}

In today's networking situations, users are mostly interested in accessing content regardless of which host is providing this content. They are looking for a fast and secure access to data in a whole range of situations: wired or wireless; heterogeneous technologies; in a fixed location or when moving. The dynamic characteristics of the network users makes the host-centric networking paradigm inefficient. Information-centric networking (ICN) is a new networking architecture where content is accessed based upon its name, and independently of the location of the hosts \cite{Zhang2010Named,Pursuit,Ahlgren2012Survey,Jacobson2009Networking}. In most ICN architectures, data is allowed to be stored in the nodes and routers within the network in addition to the content publisher's servers. This reduces the burden on the servers and on the network operator, and shortens the access time to the desired content.

Combining content routing with in-network-storage for the information is intuitively attractive, but there has been few works considering the impact of such architecture on the capacity of the network in a formal or analytical manner. In this work we study a wireless information-centric network where nodes can both route and cache content. We also assume that a node keeps a copy of the content only for a finite period of time, that is until it runs out of memory space in its cache and has to rotate content, or until it ceases to serve a specific content.

The nodes issue some queries for content that is not locally available. We suppose that there exists a server which permanently keeps all the contents. This means that the content is always provided at least by its publisher, in addition to the potential copies distributed throughout the network. Therefore, at least one replica of each content always exists in the network and if a node requests a piece of information, this data is provided either by its original server or by a cache containing the desired data. When the customer receives the content, it stores the content and shares it with the other nodes if needed.

The present paper thus investigates the asymptotic\footnote{Given two functions $f$ and $g$, we say that $f(n)=O(g(n))$ or $f(n)\preceq g(n)$ if $sup_n(f(n)/g(n))<\infty$, $f(n)=\Omega(g(n))$ or $f(n) \succeq g(n)$ if $g(n)=O(f(n))$, $f(n)=\Theta(g(n))$ or $f(n)\equiv g(n)$ if both $f(n)=O(g(n))$ and $f(n)=\Omega(g(n))$,  $f(n)=o(g(n))$ or $f(n) \prec g(n)$ if $f(n)/g(n)\rightarrow 0$, and $f(n)=\omega(g(n))$ or $f(n) \succ g(n)$ if $g(n)/f(n)\rightarrow 0$.} orders of access time and throughput capacity in such content-centric networks and addresses the following questions:

\begin{enumerate}
	\item Looking at the throughput capacity and latency, can we quantify the performance improvement brought about by a  content-centric network architecture over networks with no content sharing capability?
	\item How does the content discovery mechanism affect the performance? More specifically, does selecting the nearest copy of the content improve the scaling of the capacity and access time compared to selecting the nearest copy in the direction of original server?  
	\item How does the caching policy, and in particular, the length of time each piece of content spends in the cache's memory, affect the performance?
\end{enumerate}

We state our results in three theorems; Theorem \ref{thm:01} formulates the capacity in a grid network which uses the shortest path to the server content discovery mechanism considering different content availability in different caches, and Theorem \ref{thm:02} and \ref{thm:03} answer the above questions studying two different network models (grid and random network) and two content discovery scenarios (shortest path to the server and shortest path to the closest copy of the content) when the information exists in all caches with the same probability. These theorems demonstrate that adding the content sharing capability to the nodes can significantly increase the capacity.

The rest of the paper is organized as follows. After a brief review of the related work in Section \ref{sec:related}, the network models, the content discovery algorithms used in the current work, and the content distribution in steady-state are introduced in Section \ref{sec:netmodel}. The main theorems are stated and proved in Section \ref{sec:theorems}. We discuss the results and study some simple examples in Section \ref{sec:discussion}. Finally the paper is concluded and some possible directions for the future work will be introduced in Section \ref{sec:conclusion}.
\vspace{-0.6em}
\section{Related Work}
\label{sec:related}

Information Centric Networks have recently received considerable attention. While our work presents an analytical abstraction, it is based upon the principles described in some ICN architectures, such as CCN~\cite{Jacobson2009Networking}, NetInf~\cite{Ahlgren2008Design}, PURSUIT~\cite{Pursuit}, or DONA~\cite{Koponen2007Dataoriented}, where nodes can cache content, and requests for content can be routed to the nearest copy. Papers surveying the landscape of ICN~\cite{Ahlgren2012Survey}\cite{Ghodsi2011InformationCentric} show the dearth of theoretical results underlying these architectures. 

Caching, one of the main concepts in ICN networks, has been studied in prior works~\cite{Ahlgren2012Survey}. \cite{Olmos2014Catalog} computes the performance of a {Least-Recently-Used} (LRU) cache taking into account the dynamical nature of the content catalog. Some performance metrics like miss ratio in the cache, or the average number of hops each request travels to locate the content have been studied in \cite{InfoCom01:Che,InfoCom10:Rosensweig}, and the benefit of cooperative caching has been investigated in \cite{Wolman1999Scale}. 

Optimal cache locations \cite{Rosensweig09:Breadcrumbs}{, cach sizes \cite{Azimdoost2015Optimal}}, and cache replacement techniques \cite{IEEEMob05:Yin} are other aspects most commonly investigated. The work in \cite{Rosensweig2013Network} considers a network of LRU caches with arbitrary topology and  develops a calculus for computing bounding flows in such network. And an analytical framework for investigating properties of these networks like fairness of cache usage is proposed in \cite{Tortelli2011Fairness}. \cite{Westphal2005Maximizing} considered information being cached for a limited amount of time at each node, as we do here, but focused on flooding mechanism to locate the content, not on the capacity of the network. \cite{Dehghan2014Complexity} investigates the routing in such networks in order to minimize the average access delay. Rossi and Rossini explore the impact of multi-path routing in networks with online caching \cite{Rossi2011Caching}, and also study the performance of CCN with emphasis on the size of individual caches \cite{Rossi2012Sizing}.

However, to the best of our knowledge, there are just a few works focusing on the achievable data rates in such networks. Calculating the asymptotic throughput capacity of wireless networks with no cache has been solved in~\cite{Gupta2000Capacity} and many subsequent works~\cite{Li2001Capacity}\cite{Niesen2009Capacity}. Some work has studied the capacity of wireless networks with caching \cite{Grossglauser2002Mobility}\cite{Herdtner2005Throughput}\cite{AlfanoContentCentric} . There, caching is used to buffer data at a relay node which will physically move to deliver the content to its destination, whereas we follow the ICN assumption that caching is triggered by the node requesting the content. 
\cite{ICCW09:Liu} uses a network simulation model and evaluates the performance (file transfer delay) in a cache-and-forward system with no request for the data. \cite{Carofiglio2011Modeling} proposes an analytical model for single cache miss probability and stationary throughput in cascade and binary tree topologies. Some scaling regimes for the required link capacity is computed in \cite{Gitzenis2013Asymptotic} for a static cache placement in a multihop wireless network.

\cite{IEEEIT11:Niesen} considers a general problem of delivering content cached in a wireless network and provides some bounds on the caching capacity region from an information-theoretic point of view{, and \cite{MaddahAli2013Fundamental} proposes a coded caching scheme to achieve the order-optimal performance. Additionally, the wireless device-to-device cache networks' performance with offline caching phase has been studied in \cite{Ji2014Fundamental,Ji2016Wireless,Liu2015Asymptotic}. This is important to note that our current work is different from \cite{IEEEIT11:Niesen,MaddahAli2013Fundamental,Ji2014Fundamental,Ji2016Wireless,Liu2015Asymptotic} since unlike the mentioned works it considers the online caching and assumes that the cache contents are updated during the content delivery time.}
 
 {A preliminary version of this paper \cite{Azimdoost2013Throughput} has derived the throughput capacity when all the items have exactly the same characteristics (popularity), which has been shown to be one of the important characteristics of such networks \cite{Bharath2015LearningBased, Bastug2015Transfer}.  In this work, we do not assume any specific popularity distribution and present the results for any arbitrary request pattern.}
\vspace{-0.6em}
\section{Preliminaries}
\label{sec:netmodel}
\vspace{-0.6em}
\subsection{Network Model}
Two network models are studied in this work.

\subsubsection{Grid Network}
\begin{figure}[http]
    \center
      \includegraphics[scale=0.65,angle=0]{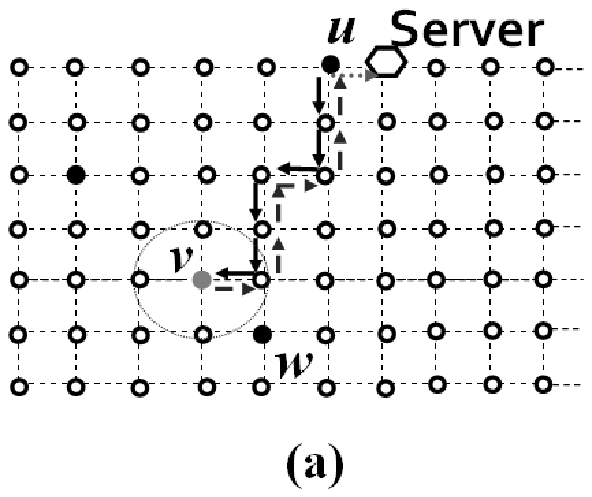} \includegraphics[scale=0.51,angle=0]{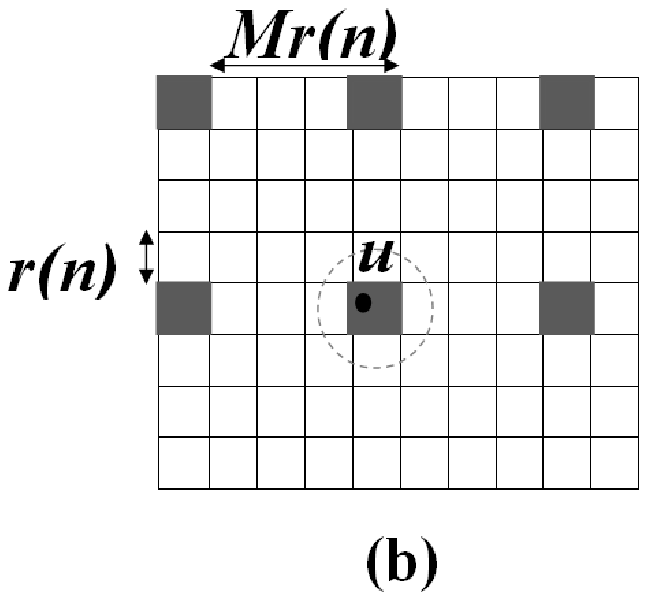} 
      \caption{a) Grid network: the transmission range of node $v$ contains four surrounding nodes. The black vertices contain the content in their  local caches. The arrow lines demonstrate a possible discovery and receive path in scenario $\mathbf{\romannumeral 1}$, where node $v$ downloads the required information from $u$. In scenario $\mathbf{\romannumeral 2}$, $v$ will download the data from $w$ instead. b) Random network:  the grey squares are the cells that can transmit simultaneously without interference, and $r(n)$ is the transmission range of each node.}
    \label{fig:gridnet}
\end{figure}

Assume that the network consists of $n$ nodes $\{v_1,v_2,...,v_n\}$ each with a local cache of size $L_i=\Theta(1)$ located on a grid. In this work we focus on the grid shown in Figure \ref{fig:gridnet}(a), but conjecture the theorems could be adapted to other regular grid topologies too. Each node can transmit over a common wireless channel, with bandwidth $W$ bits per second, shared by all nodes. The distance between two adjacent nodes equals to the transmission range of each node, so the packets sent from a node are only received by four adjacent nodes. 

There are $m$ different contents, $\{f_1,...,f_m\}$ with sizes $\{B_1,...,B_m\}$, for which each node $v_j$ may issue a query with probabilities $\{\alpha_k,\ k=1,...,m\}$, where $\sum^m_{k=1} \alpha_k=1$, and $m$ and $\alpha_k$ are not changing with the network size\footnote{In this work we are not considering applications like YouTube where the users are content generators.}. Based on the content discovery algorithms which will be explained later in this section, the query will be transmitted in the network to discover a node containing the desired content locally. $v_j$ then downloads $B_k$ bits of data with rate $\gamma$ in a hop-by-hop manner through the path $P_{xj}$ from either a node ($v_i, x=i$) containing it locally ($f\in v_i$) or the server ($x=s$). When the download is completed, the data is cached and shared with other nodes either by all the nodes on the delivery path, or only by the end node. In the paper we consider both options.  

$P_{js}$ denotes the nodes on the path from $v_j$ to server. Without loss of generality, we assume that the server is attached to the node located at the middle of the network, as changing the location of the server does not affect the scaling laws. Using the protocol model and according to \cite{book06:Xue},  the transport capacity in such network is upper bounded by $\Theta(W\sqrt{n})$. This is the model studied in Theorem \ref{thm:01} and the first two scenarios of Theorem \ref{thm:02}.

\subsubsection{Random Network}

The next network studied in Theorem \ref{thm:02} is a more general network model where the nodes are randomly distributed over a unit square area according to a uniform distribution (Figure \ref{fig:gridnet}(b)). We use the same model used in \cite{book06:Xue} (section $5$) and divide the network area into square cells each with side-length proportional to the transmission range $r(n)$, which is decreasing when the number of nodes increases, and is selected to be at least $\Theta\sqrt{\frac{\log n}{n}}$ to guarantee the connectivity of the network \cite{Applied97:Penrose} and a non-zero capacity. According to the protocol model \cite{book06:Xue}, if the cells are far enough they can transmit data at the same time with no interference; we assume that there are $M^2$ non-interfering groups which take turn to transmit at the corresponding time-slot in a round robin fashion. Again, without loss of generality the server is assumed to be located at the middle of the network. In this model the maximum number of simultaneous feasible transmissions  will be in the order of $\frac{1}{r^2(n)}$ as each transmission consumes an area proportional to $r^2(n)$. 
All other assumptions are similar to the grid network.
\vspace{-0.6em}
\subsection{Content Discovery Algorithm}
\subsubsection{Path-wise Discovery}
To discover the location of the desired content, the request is sent through the shortest path toward the server containing the requested content. If an intermediate node has the data in its local cache, it does not forward the request toward the server anymore and the requester will start downloading from the discovered cache. Otherwise, the request will go all the way toward the server and the content is obtained from the main source. In case of the random network when a node needs a piece of information, it will send a request to its neighbors toward the server, i.e. the nodes in the same cell and one adjacent cell in the path toward the server, if any copy of the data is found it will be downloaded. If not, just one node in the adjacent cell will forward the request to the next cell toward the server.
	
\subsubsection{Expanding Ring Search}
In this algorithm the request for the information is sent to all the nodes in the transmission range of the requester. If a node receiving the request contains the required data in its local cache, it notifies the requester and then downloading from the discovered cache is started. Otherwise, all the nodes that receive the request will broadcast the request to their own neighbors. This process continues until the content is discovered in a cache and the downloading follows after that. This will return the nearest copy from the requester.
\vspace{-0.6em}
\subsection{Content Distribution in Steady-State}
\label{SSanalysis}
\begin{figure}[http]
    \center
      \includegraphics[scale=0.72,angle=0]{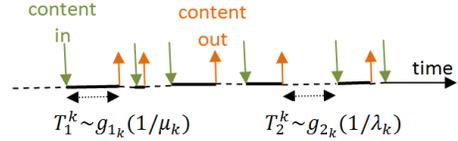}\\
      \caption{Data access process time diagram in a cache for content $k$}
    \label{fig:Tdgrm1}
\end{figure}

The time diagram of data access process in a cache is illustrated in Figure \ref{fig:Tdgrm1}.
When a query for content $f_k$ is initiated, the content is downloaded from a cache containing it and is received by another cache where it is going to be kept. The same cache may receive the same data after some random time ($T^k_2$) with distribution $g_{2_k}$ and mean $1/\lambda_k$. Note that 1) no specific caching policy is assumed here, and 2) a node will receive the content only if it does not have it in its local cache. The solid lines in this diagram denote the portions of time that the data is available at the cache.

As the requests for different contents are assumed to be independent and holding times are set for each content independent of the others, we can do the calculations for one single content. If the total number of contents is not a function of the network size, this will not change the capacity order.
Assume that content sizes $B_k$ are much larger than the request packet size, so we ignore the overhead of the discovery phase in our calculations. 

The average portion of time that each node contains a content in its local cache is
\begin{eqnarray}
\rho^{(k)}(n)=\frac{1/\mu_k}{1/\mu_k+1/\lambda_k}=\frac{\lambda_k}{\lambda_k+\mu_k}, \label{eq:rhok}
\end{eqnarray}
which is the average probability that a node contains the content $k$ at steady-state. $\lambda_k$ is the rate of requests for content $k$ received by a cache in case of the data not being available, and $\mu_k$ is the rate of the data being expunged from the cache. Both these parameters can strongly be dependent on the total number of users, or the topology and configuration of the network or the cache characteristics like size and replacement policy.
\vspace{-0.2in}
\subsection{Performance Indices}
\label{metrics}
The performance indices studied in this work are:
\subsubsection{Throughput Capacity}
 Throughput capacity is the maximum common content download rate which can be achieved by all users on average.
\subsubsection{Average Latency}
The average amount of time it takes for a customer to receive its required content from a cache or server.
\subsubsection{Total Traffic} 
The total traffic generated by downloading item $k$ is the number of item $k$ bits moving across the netwrok in a second. In other words, it is the product of total request rate (the product of the number of requesting nodes and the rate at which each node is sending the request), the number of hops between source and destination, and the content size.
\vspace{-0.6em}
\section{Theorem Statements and Proofs}
\label{sec:theorems}
Consider a grid wireless network consisting of $n$ nodes, transmitting over a common wireless channel, with shared bandwidth of $W=\Theta(1)$ bits per second. Assume that there is a server which contains all the information. Without loss of generality we assume that this server is located in the middle of the network. Each node contains some information in its local cache. Assume that according to the symmetry, the probability of each content $k$ being in all the caches with the same distance ($j$ hops) from the server is the same ($\rho^{(k)}_j(n)$).  
\begin{theorem} \label{thm:01}
The maximum achievable throughput capacity order ($\gamma_{max}$) in the above network when the nodes use the nearest copy of the required content on the shortest path toward the server is given by\footnote{Since no online caching assumption is used in this Theorem, it can be used for offline caching networks as well. However, we skip the offline results and target the networks with online caching which is the scope of this paper.}

\small \begin{eqnarray}
		&\gamma_{max}\equiv \frac{n}{\sum^m_{k=1} \alpha_k \sum_{i=1}^{\sqrt{n}} 4i \sum_{j=0}^{i-1} (i-j)\rho^{(k)}_j(n)\prod_{l=j+1}^i(1-\rho^{(k)}_l(n))},& \nonumber
\end{eqnarray}\normalsize
	
where $\rho^{(k)}_0(n)=1$, which means that the server always contains all the contents.

\end{theorem}

\begin{IEEEproof}
Considering the grid topology and large number of nodes, each cache may receive requests and downloaded contents originated from different nodes. Since the users are sending requests independent of each other, the requests received by different caches can be assumed independent of each other. Thus, the information in each cache is independent of the contents in the other caches. 
This assumption has been made in some other works too, among which are \cite{Carofiglio2011Modeling, Martina2014Unified, Fofack2012Analysis, Dabirmoghaddam2014Understanding, Psaras2011Modelling} to name a few.

A request initiated by a user $v_i$ in $i$-hop distance from the server (located in level $i=1,..,\sqrt{n}$) is served by cache $u_j$ located in level $j,\ 1\leq j\leq i$ on the shortest path from $v_i$ to the server if no caches before $u_j$, including $v_i$, on this path contains the required information, and $u_j$ contains it. This request is served by the server if no copy of it is available on the path. Let $P^{(k)}_{i,j}$ denote the probability of $v_i$'s request for item $k$ being served by $u_j$, this probability is given by $P^{(k)}_{i,j}=$
\begin{eqnarray}
&(1-\rho^{(k)}_i(n))(1-\rho^{(k)}_{i-1}(n))...(1-\rho^{(k)}_{j+1}(n))\rho^{(k)}_j(n)& \label{eq:Pk}
\end{eqnarray}
where $\rho^{(k)}_j(n)$ is the probability of content $k$ being available in a cache in level $j, 1 \leq j \leq \sqrt{n}$, and $j=0$ shows the server and $\rho^{(k)}_0(n)=1$. Thus a content $k$ requested by $v_i$ is traveling $i-j$ hops with probability $P^{(k)}_{i,j}$. There are $4i$ nodes in level $i$ so the average number of hops ($E[h_k]$) traveled by item $k$ from the serving cache (or the original server) to the requester is 
\begin{eqnarray}
&E[h_k]=\frac{1}{n}\sum_{i=1}^{\sqrt{n}} 4i \sum_{j=0}^{i-1} (i-j)P^{(k)}_{i,j}& \label{eq:barh} 
\end{eqnarray}
Therefore the average number of hops in the network is given by $E[h]=\sum^m_{k=1} \alpha_k E[h_k]$. 

Assume that each user is receiving data with rate $\gamma$. The transport capacity in this network, which equals to $n\gamma E[h]$, is upper bounded by $\Theta(\sqrt{n})$ bits-meters/sec divided by the distance of each hop $\Theta(\frac{1}{\sqrt{n}})$, which is $\Theta(n)$ bits-hops/sec. So $\gamma_{max}=\Theta(\frac{1}{E[h]})$ and the Theorem is proved.
\end{IEEEproof}

Now consider a wireless network consisting of $n$ nodes, with each node containing information $k$ in its local cache with common probability\footnote{The proof does not need the probabilities to be exactly the same, they just need to be of the same order in terms of $n$.}\vphantom{d}$^{\text{,}}$\footnote{Note that this assumption is correct for networks with online caching. In offline caching scenarios each content is present in some specific caches. However, offline caching can be considered as a special case of online caching, and we still can use this theorem by assigning the value of the fraction of caches containing the item  to the probability of each item being in a cache.}, $\rho^{(k)}(n)\nrightarrow 1$ (meaning that it does not tend to 1 when $n$ increases.), otherwise for $\rho^{(k)}(n) \rightarrow 1$, the request is served locally and no data is transferred between the nodes. Assume that the request process and cache look up time in each node is not a function of the number of nodes. Here, based on the network models and content discovery methods, we define the following different scenarios, and then  study the corresponding performance of caching in Theorems \ref{thm:02} and \ref{thm:03};
\begin{itemize}
		\item Scenario $\mathbf{\romannumeral 1}$- The nodes are located on a grid and search for the contents just on the shortest path toward the server,
		\item Scenario $\mathbf{\romannumeral 2}$- The nodes are located on a grid and use ring expansion to find contents,	
		\item Scenario $\mathbf{\romannumeral 3}$- The nodes are randomly distributed over a unit square area and use path-wise content discovery algorithm. Each node has a transmission range of $r(n)$ which at least equals to $\Theta(\sqrt{\frac{\log n}{n}})$ so the network is connected. 
\end{itemize}
\begin{theorem} \label{thm:02}
	The average latency order in the three scenarios defined above is
	
	\begin{itemize}
		\item Scenario $\mathbf{\romannumeral 1}$-   $\Theta(min(\sqrt{n},\frac{1}{\underset{k}{min}(\rho^{(k)}(n))})).$		
		\item Scenario $\mathbf{\romannumeral 2}$-  $\Theta(min(\sqrt{n},\frac{1}{\sqrt{\underset{k}{min}(\rho^{(k)}(n))}})).$			
		\item Scenario $\mathbf{\romannumeral 3}$-  $\Theta(max[1,min(\frac{1}{r(n)}, \frac{1}{\underset{k}{min}(\rho^{(k)}(n))nr^2(n)})]).$
		\end{itemize}
\end{theorem}
Here we prove Theorem \ref{thm:02} by utilizing some Lemmas. The proof of lemmas are presented in the Appendix. 
\begin{lemma} \label{lem:01} 
	Consider the wireless networks described in Theorem \ref{thm:02}. The average number of hops between the customer and the serving node (a cache or original server) for item $k$ is 
		\begin{itemize}
		\item Scenario $\mathbf{\romannumeral 1}$- $E[h_k]$ asymptotically equals to 
	\begin{flalign}
		&\frac{1}{n}\sum_{i=1}^{\sqrt{n}} i^2(1-\rho^{(k)}(n))^i+& \nonumber \\
		& \frac{\rho^{(k)}(n)}{n}\sum_{i=1}^{\sqrt{n}}i\sum_{l=1}^{i-1}l(1-\rho^{(k)}(n))^l& \label{eq:EXi}
	\end{flalign}
		\item Scenario $\mathbf{\romannumeral 2}$-  $E[h_k]$ asymptotically equals to
		
		\small\begin{flalign}
		 &\frac{1}{n}\{\sum\limits_{i=1}^{\sqrt{n}} i^2(1-\rho^{(k)}(n))^{2i^2-2i+1}+& \nonumber \\
		&\sum\limits_{i=2}^{\sqrt{n}}i\sum\limits_{l=1}^{i-1}l(1-\rho^{(k)}(n))^{2l^2-2l+1}(1-(1-\rho^{(k)}(n))^{4l})\}& \label{eq:EXii}
	  \end{flalign}\normalsize
		\item Scenario $\mathbf{\romannumeral 3}$-  $E[h_k]$ asymptotically equals to
		
		\small\begin{flalign}
		&r^2(n)\{\sum\limits_{i=2}^{\frac{1}{r(n)}} i^2(1-\rho^{(k)}(n))^{inr^2(n)}+& \nonumber \\
		&(1-(1-\rho^{(k)}(n))^{nr^2(n)})\sum\limits_{i=2}^{\frac{1}{r(n)}}i\sum\limits_{l=1}^{i-1}l(1-\rho^{(k)}(n))^{lnr^2(n)}\}& \label{eq:EXiii}
	\end{flalign}\normalsize
		\end{itemize} 
	\end{lemma}	
	\begin{lemma}\label{lem:02}
	Consider the wireless networks described in Theorem \ref{thm:02}. For sufficiently large networks, the average number of hops between the customer and the serving node (a cache or the original server) for item $k$ is 
	\begin{itemize}
		\item Scenario $\mathbf{\romannumeral 1}$- $E[h_k]$ equals $\sqrt{n}$ for $\rho^{(k)}(n) \preceq \frac{1}{\sqrt{n}}$, and $\frac{1}{\rho^{(k)}(n)}$ for $\rho^{(k)}(n) \succeq \frac{1}{\sqrt{n}}$.
		\item Scenario $\mathbf{\romannumeral 2}$- $E[h_k]$ equals $\sqrt{n}$ for $\rho^{(k)}(n) \preceq \frac{1}{n}$, and $\frac{1}{\sqrt{\rho^{(k)}(n)}}$ for $\rho^{(k)}(n) \succeq \frac{1}{n}$.
		\item Scenario $\mathbf{\romannumeral 3}$- $E[h_k]$ equals $\frac{1}{r(n)}$ for $\rho^{(k)}(n) \preceq \frac{1}{nr(n)}$, $\frac{1}{\rho^{(k)}(n) nr^2(n)}$ for $\frac{1}{nr(n)} \preceq \rho^{(k)}(n) \preceq \frac{1}{nr^2(n)}$, and $1$ for $\rho^{(k)}(n) \succeq \frac{1}{nr^2(n)}$.
		\end{itemize}
\end{lemma}
	Theorem \ref{thm:02} is now simply proved using the above Lemmas.

\begin{IEEEproof}
The average number of hops each content is traveling is $E[h]=\sum^m_{k=1} \alpha_k E[h_k]$.

We assume that the number of contents and also the popularity of each item is not changing with the network size (number of users). In the three scenarios mentioned above for the cases of  $\rho^{(k)}(n) \preceq \frac{1}{\sqrt{n}}$, $\rho^{(k)}(n) \preceq \frac{1}{n}$, and $\rho^{(k)}(n) \preceq \frac{1}{nr(n)}$, when there is at least one node with average number of hops equal to $\sqrt{n}$, $\sqrt{n}$, and $\frac{1}{r(n)}$ respectively, then that node's $E[h_k]$ in $E[h]$ defined above becomes the dominant factor.   

If $\rho^{(k)}(n) \succeq \frac{1}{\sqrt{n}}$, $\rho^{(k)}(n) \succeq \frac{1}{n}$, and $\rho^{(k)}(n) \succeq \frac{1}{nr^2(n)}$ for all the contents, in the three scenarios, respectively, then $E[h]$ in the three scenarios is given by $\sum^m_{k=1} \frac{\alpha_k}{\rho^{(k)}(n)}\equiv \frac{1}{\underset{k}{min}(\rho^{(k)}(n))}$, $\sum^m_{k=1} \frac{\alpha_k}{\sqrt{\rho^{(k)}(n)}}\equiv \frac{1}{\sqrt{\underset{k}{min}(\rho^{(k)}(n))}}$, and $\sum^m_{k=1} \alpha_k=1$.

In the third scenario, if there is no item for which $\rho^{(k)}(n) \preceq \frac{1}{nr(n)}$, but there is at least one item such that $\rho^{(k)}(n) \preceq \frac{1}{nr^2(n)}$, then $E[h]=\sum^m_{k=1} \frac{\alpha_k}{\rho^{(k)}(n)nr^2(n)} \equiv \frac{1}{\underset{k}{min}(\rho^{(k)}(n)nr^2(n))}$.

The total $E[h]$ can be simply written as the results shown in \textit{Theorem \ref{thm:02}}.  

Assuming that the delay of the request process and cache look up in each node is not increasing when the network size (the number of nodes) increases, and that there is enough bandwidth to avoid congestion, then the latency of getting the data is directly proportional to the average number of hops between the serving node and the customer. Thus, the latency and the average number of hops the data is traveling to reach the customer are of the same order and \textit{Theorem \ref{thm:02}} is proved.
\end{IEEEproof}

\begin{theorem}\label{thm:03}
Consider the networks of Theorem \ref{thm:02}, and assume each node can transmit over a common wireless channel, with $W=\Theta(1)$ bits per second bandwidth, shared by all nodes. The maximum achievable throughput capacity order $\gamma_{max}$  in the three discussed scenarios are  
        \begin{itemize}
		\item Scenario $\mathbf{\romannumeral 1}$- $\Theta(max(\frac{1}{n}, \underset{k}{min}((\rho^{(k)}(n))^2))).$
		
		\item Scenario $\mathbf{\romannumeral 2}$- $\Theta(max(\frac{1}{n}, \underset{k}{min}(\rho^{(k)}(n)))).$	
		
		\item Scenario $\mathbf{\romannumeral 3}$-	\\ $\Theta(max[\frac{1}{n},min(\frac{1}{nr^2(n)},\underset{k}{min}((\rho^{(k)}(n))^2)nr^2(n))]).$
		\end{itemize}
		
\end{theorem}
To prove Theorem \ref{thm:03} we use Lemma \ref{lem:02}, and the following two Lemmas.
\begin{lemma}\label{lem:03}
	Consider the wireless networks described in Theorem \ref{thm:02}. In order not to have interference, the maximum throughput capacity is upper limited by  
		\begin{itemize}
		\item Scenario $\mathbf{\romannumeral 1}$- $\Theta(max(\frac{1}{\sqrt{n}},\underset{k}{min}(\rho^{(k)}(n)))).$
		\item Scenario $\mathbf{\romannumeral 2}$- $\Theta(max(\frac{1}{\sqrt{n}},\sqrt{\underset{k}{min}(\rho^{(k)}(n))})).$
		\item Scenario $\mathbf{\romannumeral 3}$- \small $\Theta(min[\frac{1}{nr^2(n)},max(\frac{1}{nr(n)}, \underset{k}{min}(\rho^{(k)}(n)))]).$ \normalsize
		\end{itemize}
\end{lemma}	
		In the previous Lemma, the maximum throughput capacity in a wireless network utilizing caches has been calculated such that no interference occurs. Now it is important to verify if this throughput can be supported by each node (cell), i.e. the traffic carried by each node (cell) is not more than what it can support ($\Theta(1)$).
\begin{lemma} \label{lem:04}
The maximum supportable throughput capacities in the studied scenarios are as follows.
        
	\begin{itemize}
		\item Scenario $\mathbf{\romannumeral 1}$-		$\Theta(max(\frac{1}{n}, \underset{k}{min}((\rho^{(k)}(n))^2)))$.
		\item Scenario $\mathbf{\romannumeral 2}$- $\Theta(max(\frac{1}{n}, \underset{k}{min}(\rho^{(k)}(n))))$.
		\item Scenario $\mathbf{\romannumeral 3}$- \\ $\Theta(max[\frac{1}{n},min(\frac{1}{nr^2(n)},\underset{k}{min}((\rho^{(k)}(n))^2)nr^2(n))])$.
		\end{itemize}
	\end{lemma}
The maximum throughput capacity is the value which can be supported by all the nodes while no interference occurs.  Thus the throughput capacity will be the minimum of the two values derived in Lemmas \ref{lem:03} and \ref{lem:04}, and Theorem \ref{thm:03} is proved.
\vspace{-0.6em}
\section{Discussion}
\label{sec:discussion}
The Theorems above express the maximum achievable data download rate in terms of the availability of the contents in the caches($\rho^{(k)}(n)$), in networks with specific topology and content discovery mechanisms. However, no assumption on the caching policy, which is an important factor in determining $\rho^{(k)}(n)$ have been made. In this section, we discuss our results based on two examples and try to study the affect of caching policy on the performance. 

In these examples we consider two different cache replacement policies based on Time-To-Live (TTL). First example uses exponentially distributed TTL, and the second one considers constant TTL. According to \cite{Fofack2014Performance} this can   predict metrics of interest on networks of caches running other replacement algorithms like LRU, FIFO, or Random. 

In order to use the stated theorems, the probability of each item being in each cache is first calculated, and then, the appropriate theorem is used to give the throughput capacity. In the first example, in addition to the capacity, we analyze the total request rate ($n(1-\rho^{(k)})\lambda_k$) and total generated traffic for an item $k$ ($n(1-\rho^{(k)})\lambda_k B_k E[h_k]$) as well. This gives us an idea about how the request rates and cache holding times affect the traffic in the network and how the resources are utilized.

\vspace{-0.2in} 
\subsection{Example 1}
\label{ex:01}
\subsubsection{Network Model}
Consider a network where the received data is stored only at the receivers (edge caching \cite{Bastug2014Living,Golrezaei2012FemtoCaching}) and then shared with the other nodes as long as the node keeps the content. Assume that receiving a data $k$ in the local cache of the requesting user sets a time-out timer with exponentially distributed duration with parameter $\eta_k$ and no other event will change the timer until it times-out, meaning that in equation \eqref{eq:rhok} $\mu_k=\eta_k$, which is not a function of $n$. Considering the request process for each content $k$ from each user being a Poisson process with rate $\beta_k$ not changing with $n$, and using the memoryless property of exponential distribution (internal request inter-arrival times), and assuming that the received data is stored only in the end user's cache (the caches on the download path do not store the downloading data), it can be proved that in equation \eqref{eq:rhok} $\lambda_k=\beta_k$. Thus we can write the presence probability of each content $k$ in each cache as $\rho^{(k)}(n)=\frac{\beta_k}{\beta_k+\eta_k}$ (equal order probability of all the caches containing an item $k$). 
\subsubsection{Results}
\begin{figure}[http]
    \center
			\includegraphics[scale=0.24,angle=0]{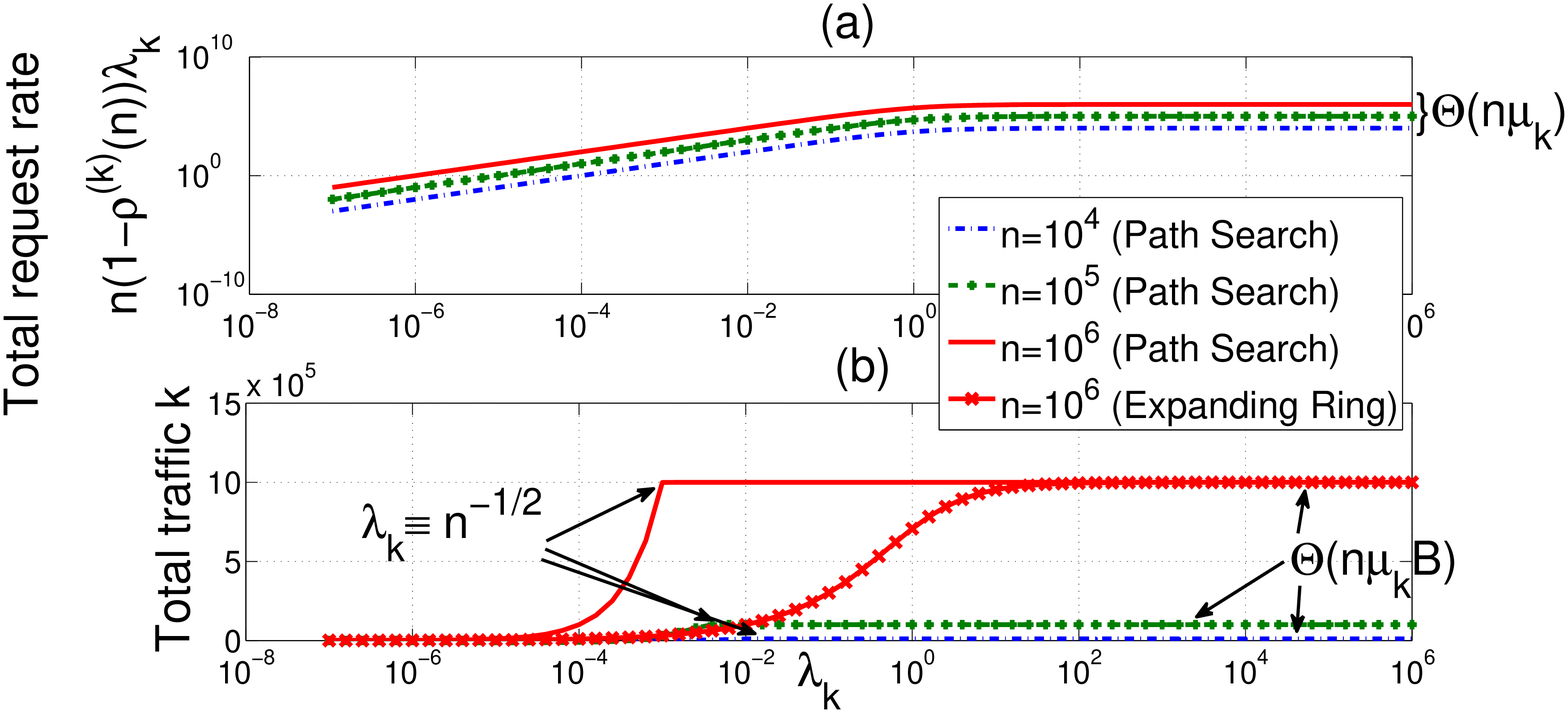}\\
      \caption{Scenario $\mathbf{\romannumeral 1}$ (a) Total request rate for an item $k$ in the network ($\lambda_k n (1-\rho^{(k)}(n))$), (b) Total traffic in the network ($B_k\lambda_k n (1-\rho^{(k)}(n))E[h_k]$) vs. the request rate ($\lambda_k$) with fixed time-out rate ($\mu_k=1$).}
    \label{fig:traffic_lambda}
\end{figure}

Figures \ref{fig:traffic_lambda} (a),(b) respectively illustrate the total request rate and the total traffic generated in a fixed size network in scenario $\mathbf{\romannumeral 1}$ for each item $k$ for different request rates when the time-out rate is fixed. 
Small $\lambda_k$ means that each node is sending requests for $k$ with low rate, so fewer caches have that content, and consequently more nodes are sending requests with this low rate. In this case most of the requests are served by the server. The total request rate of item $k$ will increase by increasing the per node request rate. High $\lambda_k$ shows that each node is requesting the content with higher rate, so the number of cached content $k$ in the network is high, thus fewer nodes are requesting it with this high rate externally. Here most of the requests are served by the caches. The total request rate then is determined by the content drop rate. So for very large $\lambda_k$, the total request rate is the total number of nodes in the network times the drop rate ($n\mu_k$) and the total traffic is $n\mu_k B_k$. As can be seen there is some request rate at which the traffic reaches its maximum; this happens when there is a balance between the requests served by the server and by the caches. For smaller request rates, most of the requests are served by the server and increasing $\lambda_k$ increases the total traffic. For larger $\lambda_k$, on the other hand, most of the requests are served by the caches and increasing the request rate will not change the distance to the nearest content and the total traffic.

\begin{figure}[http]
    \center
			\includegraphics[scale=0.24,angle=0]{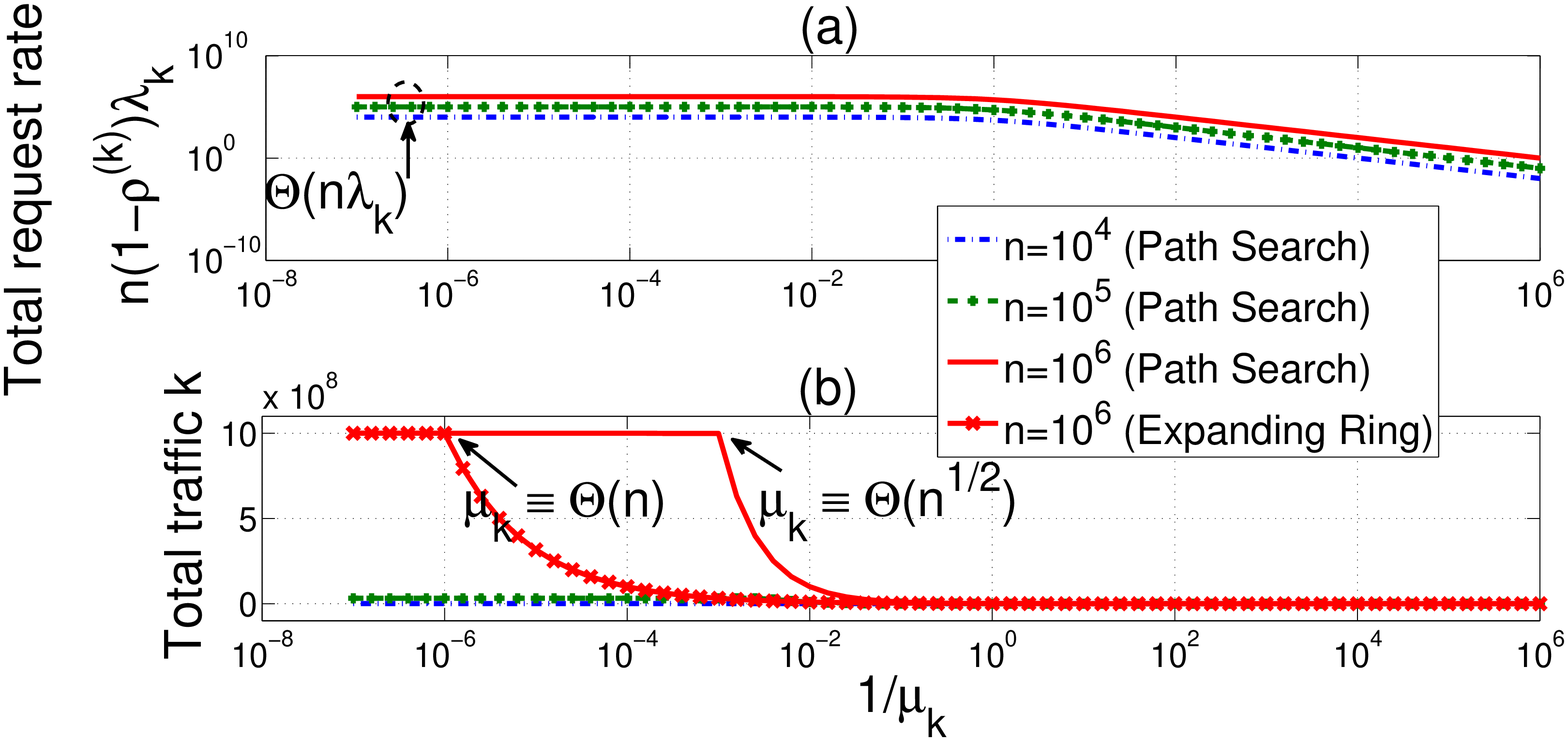}\\
      \caption{Scenario $\mathbf{\romannumeral 1}$ (a) Total request rate in the network ($\lambda_k n (1-\rho^{(k)}(n))$), (b) Total traffic in the network ($B_k\lambda_k n (1-\rho^{(k)}(n))E[h_k]$) vs. the inverse of the time-out rate ($1/\mu_k$) with fixed request ratio ($\lambda_k=1$).}
    \label{fig:traffic_mu}
\end{figure}

Figures \ref{fig:traffic_mu} (a),(b) respectively illustrate the total request rate and the total traffic generated in a fixed size network in scenario $\mathbf{\romannumeral 1}$ for different time-out rates when the request rate is fixed. For low $1/\mu_k$ (high time-out rates or small lifetimes), most of the item $k$ requests are served by the server and caching is not used at all. For large time-out times, all the requests are served by the caches, and the only parameter in determining the total request rate is the time-out rate.

However, when the network grows the traffic in the network will increase and the download rate will decrease. If we assume that the new requests are not issued in the middle of the previous download, the request rate will decrease with network growth. If the holding time of the contents in a cache increases accordingly the total traffic will not change, i.e. if by increasing the network size the requests are issued not as fast as before, and the contents are kept in the caches for longer times, the network will perform similarly.

\begin{figure}[http]
    \center
			\includegraphics[scale=0.23,angle=0]{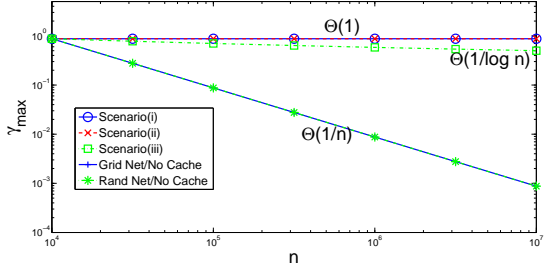}\\
      \caption{Maximum download rate ($\gamma_{max}$) vs. the number of nodes ($n$) for $\rho=7/8$.}
    \label{fig:capacity}
\end{figure}

In Figure \ref{fig:capacity} we assume that the request rate is roughly $7$ times the drop rate for all the contents, so $\rho^{(k)}(n)=7/8$, and show the maximum throughput order as a function of the network size. In scenario $\mathbf{\romannumeral 3}$, we set the transmission range to the minimum value needed to have a connected network ($r(n)\equiv \sqrt{\frac{\log n}{n}}$). According to Theorem \ref{thm:03} and as can be observed from this figure, the maximum throughput capacity of the network in a grid network with the described characteristics is not changing with the network size if the probability of each item being in each cache is fixed, while in a network with no cache this capacity will be inversely proportional to the network size. Similarly in the random network the maximum throughput is inversely proportional to $nr^2(n)$, which is the logarithm of the network size, while in a no cache network it is diminishing with the rate of network growth. 

Moreover, comparing scenario $\mathbf{\romannumeral 1}$ with $\mathbf{\romannumeral 2}$, we observe that the throughput capacity in both cases are almost the same; meaning that using the path discovery scheme will lead to almost the same throughput capacity as the expanding ring discovery. Thus, we can conclude that just by knowing the address of a server containing the required data and forwarding the requests through the shortest path toward that server we can achieve the best performance, and increasing the complexity and control traffic to discover the closest copy of the required content does not add much to the capacity. 

On the other hand with a fixed network size, if the probability of an item being in each cache is greater than a threshold ($\Theta(\frac{1}{\sqrt{n}})$, $\Theta(\frac{1}{n})$, and $\Theta(\frac{1}{nr^2(n)})=\Theta(\frac{1}{\log n})$ in cases $\mathbf{\romannumeral 1},\mathbf{\romannumeral 2}$ and $\mathbf{\romannumeral 3}$, respectively), most of the requests will be served by the caches and not the server, so increasing the probability of an intermediate cache having the content reduces the number of hops needed to forward the content to the customer, and consequently increases the throughput. For content presence probability orders less than these thresholds ($\Theta(\frac{1}{nr(n)})=\Theta(\frac{1}{\sqrt{n\log n}})$ in cases $\mathbf{\romannumeral 3}$) most of the requests are served by the main server, so the maximum possible number of hops will be traveled by each content to reach the requester and the minimum throughput capacity ($\Theta(\frac{1}{n})$) will be achieved. Note that in these networks, the maximum throughput is limited by the maximum supportable load on each link, and more specifically on the server.

As may have been expected and according to our results, the obtained throughput is a function of the probability of each content being available in each cache, which in turn is strongly dependent on the network configuration and cache management policy. 
\vspace{-0.2in}
\subsection{Example 2}
\label{ex:02}
\subsubsection{Network Model}
Assume an $n$-cache grid wireless network with one server containing all the items located in the middle of the network.  Each cache in level $i$ (nodes at $i$ hops away from the server) receives requests for a specific document $k$ according to a Poisson distribution with rate $\beta^{(k)}$ from the local user, and with rate $\beta^{(k)'}_i(n)$ from all the other nodes. Note that rate $\beta^{(k)'}_i(n)$ is a function of the individual request rate of users for item $k$ ($\beta^{(k)}$) and also the location of the cache in the network. The content discovery mechanism is path-wise discovery, and whenever a copy of the required data is found (in a cache or server), it will be downloaded through the reverse path, and either all the nodes on the download path or only the requester node store it in their local caches. Moreover, we assume that receiving the item $k$ and also any request for the available cached data $k$ by a node in level $i$ refreshes a time-out timer with fixed duration $D^{(k)}_i(n)$. According to \cite{Che2002Hierarchical},  this is a good approximation for caches with LRU replacement policy when the cache size and the total number of documents are reasonably large. Furthermore, according to the same work this value is a constant for all contents and is a function of the cache size, so we can use $D_i(n)$ for all contents in caches in level $i$. We will calculate the average probability of item $k$ being in a cache in level $i$ ($\rho^{(k)}_i(n)$) based on these assumptions and then use Theorem \ref{thm:01} to obtain the throughput capacity.

\subsubsection{Results}
Let random variable $t^{(k)i}_{on}(T)$ denote the total time of the data $k$ being available in a cache in level $i$ ($i$ hop distance from the server) during constant time $T$. Assume that item $k$ is received $N^{(k)i}(T)$ times during time $T$ by each node $v_i$ in level $i$ (according to the symmetry all nodes in one level have similar conditions.). The data available time between any two successive receipt of item $k$ is $D_i(n)$ if the timer set by the first receipt is expired before the second one comes, or is equal to the time between these two receipts. Let $\tau^{req(k)}_i$ denote the time between two successive receipts. This process has an exponential distribution with parameter $\beta^{(k)}_i=\beta^{(k)}+\beta^{(k)'}_i$. So the total time of data $k$ availability in a level $i$ cache is 
\begin{eqnarray}
t^{(k)i}_{on}(T)=\sum_{j=0}^{N^{(k)i}(T)} min(\tau_i^{req(k)},D_i(n)),
\end{eqnarray} 
and the average value of this time is ($E[t^{(k)i}_{on}(T)]$) 

\begin{eqnarray}
&&\sum_{l=0}^{\infty} E[\sum_{j=0}^{l} min(\tau_i^{req(k)},D_i(n))]Pr(N^{(k)i}(T)=l), \nonumber \\
&=&\sum_{l=0}^{\infty} lE[min(\tau_i^{req(k)},D_i(n))]Pr(N^{(k)i}(T)=l), \nonumber \\
&=&E[min(\tau_i^{req(k)},D_i(n))]E[N^{(k)i}(T)]. \label{eq:Eton}
\end{eqnarray}

According to the Poisson arrivals of requests (data downloads) with parameter $\beta^{(k)}+\beta^{(k)'}_i$, the rightmost term in equation \eqref{eq:Eton} ($E[N^{(k)i}(T)]$) equals $(\beta^{(k)}+\beta^{(k)'}_i)T$. The leftmost term in this equation ($E[min(\tau_i^{req(k)},D_i(n))]$) can also be easily calculated and equals to  $\frac{1-e^{-D_i(n)(\beta^{(k)}+\beta^{(k)'}_i)}}{\beta^{(k)}+\beta^{(k)'}_i}$. Therefore, $E[t^{(k)i}_{on}(T)]=(1-e^{-D_i(n)(\beta^{(k)}+\beta^{(k)'}_i)})T$. And finally the probability of an item $k$ being available in a level $i$ cache is $\rho^{(k)}_i=\frac{E[t^{(k)i}_{on}(T)]}{T}=1-e^{-D_i(n)(\beta^{(k)}+\beta^{(k)'}_i(n))}$. 
Note that $D_0=\infty$ so that $\rho^{(k)}_0=1$.

Now we need to calculate the rate of item $k$ received by each node in level $i$. First, assume that when an item is downloaded , only the end user (the node which has requested the content) keeps the downloaded content, and storing a new content refreshes the time-out timer with fixed duration $D_i(n)$. Thus $\beta^{(k)'}_i(n)=0$, and $\rho^{(k)}_i(n)=1-e^{-D_i(n)\beta^{(k)}}$. It is obvious that in such network where all the caches have the same size and the request patterns, $D_i(n)$ will not depend on the cache location, and since the request rate and the caches sizes are not changing with $n$ this value does not depend on the network size either. Thus, $D_i(n)$ can be replaced by fixed and constant $D$. Therefore, $\rho^{(k)}_i(n)=1-e^{-D\beta^{(k)}}$ which is similar for all the caches, and the maximum throughput capacity order ($\gamma_{max}$) is $\frac{n}{\sum_{k=1}^m \alpha_k\sum_{i=1}^{\sqrt{n}}i\sum_{j=0}^{i-1}(i-j)(1-e^{-D\beta^{(k)}})e^{-(i-j)D\beta^{(k)}}}$, which is
\begin{eqnarray}
  \frac{1}{\sum_{k=1}^m \frac{\alpha_k e^{-D\beta^{(k)}}}{1-e^{-D\beta^{(k)}}}}\equiv 1. 
\end{eqnarray}

As the second case, we assume that all the nodes on the download path keep the data, and the shortest path from the requester to the server is selected such that all the nodes in level $i$ receive the requests for item $k$ with the same rate. There are $4i$ nodes in level $i$ and $4(i+1)$ nodes in level $i+1$. So the request initiated or forwarded from a node in level $i+1$ will be received by a specific node in level $i$ with probability $\frac{i}{i+1}$ if it is not locally available in that node, so $\beta^{(k)'}_i(n)$ can be expressed as
\begin{eqnarray}
\beta^{(k)'}_i=\frac{(1-\rho^{(k)}_{i+1})(\beta^{(k)}+\beta^{(k)'}_{i+1})(i+1)}{i}  \label{eq:beta}
\end{eqnarray} 

Combining equation \eqref{eq:beta}, the relationship between $\rho^{(k)}_i$ and $\beta^{(k)'}_i$, and the fact that there is no external request coming to the nodes at the edge boundary of the network ($\beta^{(k)'}_{\sqrt{n}}=0$), together with the result of Theorem \ref{thm:01} we can obtain the capacity ($\gamma_{max}$) in the grid network with path-wise content discovery and on-path storing scheme which is $n$ divided by $\sum_{k=1}^m \alpha_k \sum_{i=1}^{\sqrt{n}}i\sum_{j=0}^{i-1} (i-j)(1-e^{-D_j(n)(\beta^{(k)}+\beta^{(k)'}_j)})e^{-\sum_{l=j+1}^i D_l(n)(\beta^{(k)}+\beta^{(k)'}_l)}$.  

The result of this equation cannot exceed $\Theta(1)$ since this is the maximum possible throughput order in the grid network. Thus, caching the downloaded data in all the caches on the download path does not add any asymptotic benefit in the capacity of the network, and keeping the downloaded items only in the requester caches will yield the maximum possible throughput.
\vspace{-0.6em}
\section{Conclusion And Future Work}
\label{sec:conclusion}
We studied the asymptotic throughput capacity and latency of ICNs with limited lifetime cached data at each node. The grid and random networks are two network models we investigated in this work. Representing all the results in terms of the probability of the items being in the caches while not considering any specific content popularity distribution, or cache replacement policy has empowered us to have a generalized result which can be used in different scenarios. Our results show that with fixed content presence probability in each cache, the network can have the maximum throughput order of $1$ and $\frac{1}{nr^2(n)}$ in cases of grid and random networks, respectively, and the number of hops traveled by each data to reach the customer (or latency of obtaining data), can be as small as one hop. 

Moreover, we studied the impact of the content discovery mechanism on the performance in grid networks. It can be observed that looking for the closest cache containing the content will not have much asymptotic advantage over the simple path-wise discovery when $\underset{k}{min} \rho^{(k)}(n)$ is sufficiently small ($\underset{k}{min} \rho^{(k)}(n) \preceq \frac{1}{n}$) or big enough ($\underset{k}{min} \rho^{(k)}(n) \nrightarrow 0$). For other values of $\underset{k}{min} \rho^{(k)}(n)$, looking for the nearest copy at most decreases the throughput diminishing rate by a factor of two. Consequently, downloading the nearest available copy on the path toward the server has similar performance as downloading from the nearest copy. A practical consequence of this result is that routing may not need to be updated with knowledge of local copies, just getting to the source and finding the content opportunistically will yield the same benefit. 

Another interesting finding is that whether all the caches on the download path keep the data or just the end user does it, the maximum throughput capacity scale does not change.

In this work, we represented the fundamental limits of caching in the studied networks, proposing a caching and downloading scheme that can improve the capacity order is part of our future work.
\vspace{-0.6em}
\appendix

\begin{IEEEproof}[Proof of Lemma \ref{lem:01}]
Let $h_k$, $d_{sr}$, and $d_{max}$ denote the number of hops between the customer and the serving node (cache or original server) for content $k$, the number of hops between the customer and the serving node (cache or original server), and the maximum value of $d_{sr}$, respectively. The average number of hops between the customer and the serving node ($E[h_k]$) is given by
\begin{eqnarray}
E[h_k]=\sum_{i=1}^{d_{max}} E[h_k|d_{sr}=i]Pr(d_{sr}=i). \label{eq:hbar}
\end{eqnarray}

Scenario $\mathbf{\romannumeral 1}$- This case can be considered as a special case of the network studied in Theorem \ref{thm:01}, where $\rho^{(k)}_i(n)$ is the same for all $i$\footnote{We will give examples in Section V using this assumption.}. Thus, we can drop the index $i$ and let $\rho^{(k)}(n)$ denote the common value of this probability. Using equation \eqref{eq:Pk} and \eqref{eq:barh} we will have $E[h_k]$ equal to

\small\begin{equation}
\frac{4}{n}\sum_{i=1}^{\sqrt{n}} i \{i(1-\rho^{(k)}(n))^i+\sum_{j=1}^{i-1} (i-j)(1-\rho^{(k)}(n))^{i-j}\rho^{(k)}(n)\} 
\end{equation}

\normalsize The constant factor $4$ does not change the scaling order and it can be dropped. By defining $l=i-j$, then the proof follows.

Scenario $\mathbf{\romannumeral 2}$ - $d_{max}$ in this network is $\Theta(\sqrt{n})$, and there are $4i$ nodes at distance of $i$ hops from the original server. Thus, $Pr(d_{sr}=i)\equiv\frac{i}{n}$.
Each customer may have the required item $k$ in its local cache with probability $\rho^{(k)}(n)$. If the requester is one hop away from the original server, it gets the required item from the server with probability $1-\rho^{(k)}(n)$. The customers at two hops distance from the server ($8$ such customers) download the required item from the original server (traveling $h_k=2$ hops) if no cache in a diamond of two hops diagonals contains it (with probability $(1-\rho^{(k)}(n))^2$), and gets it from a cache at distance one hop if one of those caches has the item (with probability $(1-\rho^{(k)}(n))(1-(1-\rho^{(k)}(n))^4)$). Using similar reasoning, the customers at distance $i$ from the server get the item from the server (distance $h_k=i$ hops) with probability $(1-\rho^{(k)}(n))^{1+4(1+2+...+(i-1))}=(1-\rho^{(k)}(n))^{2i^2-2i+1}$, and from a cache at distance $h_k=l<i$ with probability $(1-\rho^{(k)}(n))^{2l^2-2l+1}(1-(1-\rho^{(k)}(n))^{4l})$ as there are $4l$ nodes at distance of $l$ hops. Therefore, using equations \eqref{eq:hbar}, \eqref{eq:Pk}, and \eqref{eq:barh}

\footnotesize\begin{flalign}
& E[h_k]\equiv \frac{1}{n}\sum_{i=2}^{\sqrt{n}} i\sum_{l=1}^{i-1} l(1-(1-\rho^{(k)}(n))^{4l})(1-\rho^{(k)}(n))^{2l^2-2l+1}& \nonumber \\
&+\frac{1}{n}\sum_{i=1}^{\sqrt{n}} i^2(1-\rho^{(k)}(n))^{2i^2-2i+1}&
\end{flalign}\normalsize

Scenario $\mathbf{\romannumeral 3}$ -  The number of caches within transmission range (one hop) is $\Theta(nr^2(n))$. $d_{max}$ in this network is of the order of $\frac{1}{r(n)}$ and $\Pr(d_{sr}=i) \equiv ir^2(n)$.

Each customer may have the required item $k$ in its local cache with probability $\rho^{(k)}(n)$.  If the requester is one hop away from the original server ($4\Theta(nr^2(n))$ nodes), it receives the required item from the server with probability $1-\rho^{(k)}(n)$. The customers at two hops distance from the server ($8\Theta(nr^2(n))$ such customers) download the required item from the original server (traveling $h_k=2$ hops) if no cache in the cell at one hop distance contains it (probability $(1-\rho^{(k)}(n))^{2nr^2(n)}$), and gets it from a cache at distance one hop if one of those caches has the item (probability $(1-\rho^{(k)}(n))(1-(1-\rho^{(k)}(n))^{2nr^2(n)})$). Using similar reasoning the customers at distance $i$ from the server receive the item from the server (distance $h_k=i$ hops) with probability $(1-\rho^{(k)}(n))^{inr^2(n)}$, and from a cache at distance $h_k=l<i$ with probability $(1-\rho^{(k)}(n))^{lnr^2(n)}(1-(1-\rho^{(k)}(n))^{nr^2(n)})$. Therefore, according to equation \eqref{eq:hbar} $E[h_k]$ equals to

\small\begin{equation}
	 r^2(n)\{(1-\rho^{(k)}(n))+\sum_{i=2}^{\frac{1}{r(n)}} i^2(1-\rho^{(k)}(n))^{inr^2(n)} \nonumber 
	 \end{equation}
	 \begin{eqnarray}
	+(1-(1-\rho^{(k)}(n))^{nr^2(n)})\sum_{i=2}^{\frac{1}{r(n)}}i\sum_{l=1}^{i-1}l(1-\rho^{(k)}(n))^{lnr^2(n)}\}. \nonumber \\
	\end{eqnarray}\normalsize

	Noting that $r^2(n)(1-\rho^{(k)}(n))$ is always less than one, and tends to zero for sufficiently large $n$, the Lemma is proved.
	\end{IEEEproof}
	 
	\begin{IEEEproof}[Proof of Lemma \ref{lem:02}]
To simplify the notations, we have dropped the index $k$ when there is no ambiguity. 

To prove this Lemma we use (A): $\lim_{N\rightarrow \infty} (1-x)^N \approx e^{-xN}$ approximation\label{eq:e_xn}, which is $\approx 1$ for $x=o(\frac{1}{N})$ (region 1), $\approx e^{-1}$ for $x=\Theta(\frac{1}{N})$ (region 2), and $\approx 0$ for $x=\omega(\frac{1}{N})$ (region 3).

Scenario $\mathbf{\romannumeral 1}$ - Let us define 

\small\begin{equation}
E_s^i=\frac{1}{n}\sum_{i=1}^{\sqrt{n}} i^2(1-\rho(n))^i  \ \ \ , \ \ \ 
E_c^i=\frac{\rho(n)}{n}\sum_{i=1}^{\sqrt{n}}i\sum_{l=1}^{i-1}l(1-\rho(n))^l.
\end{equation}\normalsize

Thus equation \eqref{eq:EXi} is written as $E[h]=E_s^i+E_c^i$.
First we investigate the value of $E_s^i$ for different ranges of $\rho(n)$. The summation for $E_s^i$ can be decomposed into two  summations.

\small\begin{eqnarray}
E_s^i&\equiv& \frac{1}{n}\{\sum_{i\prec \sqrt{n}}i^2(1-\rho(n))^i + \sum_{i\equiv \sqrt{n}}i^2(1-\rho(n))^i\}
\end{eqnarray}\normalsize

Assume $\rho(n)\equiv \frac{1}{\sqrt{n}}$, then using first and second region of equation \eqref{eq:e_xn} we have
\begin{eqnarray}
E_s^i\equiv \frac{1}{n}\{\sum_{i\prec \sqrt{n}}i^2 + \sum_{i\equiv \sqrt{n}}i^2\} \equiv \frac{n^{3/2}}{n} \equiv \sqrt{n}.
\end{eqnarray}

Moreover it can easily be seen that $E_s^i$ is a decreasing function of $\rho(n)$, so for $\rho(n)$ with order less than $\frac{1}{\sqrt{n}}$ it is more than $\sqrt{n}$. Since $d_{max}=\sqrt{n}$, we can say $E_s^i\equiv \sqrt{n}$ for $\rho(n) \preceq \frac{1}{\sqrt{n}}$. Now we expand the summation to obtain 

\small\begin{flalign}
&E_s^i \equiv \frac{(1-\rho(n))(2-\rho(n))}{n\rho^3(n)}-\frac{(1-\rho(n))^{\sqrt{n}+1}}{n\rho^3(n)}& \nonumber \\
&\times\{n(1-\rho(n))^2-(1-\rho(n))(2n+2\sqrt{n}-1)+(\sqrt{n}+1)^2\}&
\end{flalign}\normalsize

If $\rho(n) \succ \frac{1}{\sqrt{n}}$, then using third region in equation \eqref{eq:e_xn}, $(1-\rho(n))^{\sqrt{n}+1}$ is going to zero exponentially, so $n(1-\rho(n))^{\sqrt{n}+1}\rightarrow 0$. Thus, $E_s^i \equiv \frac{1}{n\rho^3(n)}$, and in summary
\begin{eqnarray}
E_s^i \equiv \left\{
	\begin{array}{ll}
					\sqrt{n}& \ \ \  \rho(n)\preceq \frac{1}{\sqrt{n}} \\
					\frac{1}{n\rho^3(n)}& \ \ \  \rho(n)\succ \frac{1}{\sqrt{n}}
			\end{array}
			\right . \label{eq:Esi}
\end{eqnarray}

According to equation \eqref{eq:Esi} and since $E[h]=E_s^i+E_c^i$, when $E_s^i\equiv \sqrt{n}$ (for $\rho(n)\preceq \frac{1}{\sqrt{n}}$) which is the maximum possible order for $E[h]$, then adding $E_s^i$ to $E[h]$ cannot increase its order beyond the maximum possible value. Now to derive the order of $E[h]$ for other values of $\rho(n)$, we decompose the equation of $E_c^i= E_c^{i1}+E_c^{i2}$ to the following summations and investigate their behaviors when $\rho(n)\succ \frac{1}{\sqrt{n}}$.
\begin{eqnarray}
&E_c^{i1}=\frac{1}{n}\sum_{i\equiv \sqrt{n}}i\sum_{l=1}^{i-1}l\rho(n)(1-\rho(n))^l,& \nonumber \\
&E_c^{i2}=\frac{1}{n}\sum_{i\prec \sqrt{n}}i\sum_{l=1}^{i-1}l\rho(n)(1-\rho(n))^l&
\end{eqnarray}

The number of $i\equiv \sqrt{n}$ is in the order of $\Theta(1)$. Therefore using the following series $\sum_{x=1}^{n} x a^x$ $= \frac{a^{n+1}(n a - n - 1)+a}{(a-1)^2}$, we have $E_c^{i1}\equiv \frac{1}{\sqrt{n}}\sum_{l=1}^{\sqrt{n}}l\rho(n)(1-\rho(n))^l
\equiv \frac{1-\rho(n)}{\rho(n)\sqrt{n}}(1-(1-\rho(n))^{\sqrt{n}}(1+\rho(n)\sqrt{n}))$,
which is equivalent to $\frac{1}{\rho(n)\sqrt{n}}$ when $\rho(n)\succ \frac{1}{\sqrt{n}}$.

Utilizing the same series, the first summation in $E_c^{i2}$ is $\Theta(\sqrt{n})$. Hence we arrive at

\small\begin{eqnarray}
&E_c^{i2}\equiv \frac{1-\rho(n)}{\rho(n)n}\sum\limits_{i\prec \sqrt{n}} i[1-\{1-\rho(n)+\rho(n)i\}(1-\rho(n))^{i-1}]& \nonumber \\
&\equiv \frac{1-\rho(n) \{1- \frac{1}{n}\sum\limits_{i\prec \sqrt{n}} i(1-\rho(n))^i - \frac{1}{n}\sum\limits_{i\prec \sqrt{n}} i^2\rho(n)(1-\rho(n))^{i-1} \}}{\rho(n)}& \nonumber \\
&\equiv \frac{1-\rho(n)}{\rho(n)}-\frac{(1-\rho(n))^2}{\rho^3(n)n}-\frac{1}{\rho^3(n)n} \equiv \frac{1}{\rho(n)}&
\end{eqnarray}\normalsize

Since  $\rho(n)\succ \frac{1}{\sqrt{n}}$, $E_c^{i2}$ is the dominant factor in $E_c^i$, and also it is dominant factor in $E[h]$. Thus, $E[h]\equiv E_s^i \equiv \sqrt{n}$ for $\rho(n)\preceq \frac{1}{\sqrt{n}}$, and $E[h]\equiv E_c^{i2}\equiv \frac{1}{ \sqrt{\rho(n)}}$ for $\rho(n)\succ \frac{1}{\sqrt{n}}$.

Scenario $\mathbf{\romannumeral 2}$ - Let us define 

\small\begin{flalign}
&E_s^{ii}= \frac{1}{n}\sum_{i=1}^{\sqrt{n}} i^2(1-\rho(n))^{2i^2-2i+1},& \nonumber \\
&E_c^{ii}= \frac{1}{n}\sum_{i=2}^{\sqrt{n}}i\sum_{k=1}^{i-1}l(1-\rho(n))^{2l^2-2l+1}(1-(1-\rho(n))^{4l})&
\end{flalign}\normalsize

So $E[h]= E_s^{ii}+E_c^{ii}$. Assume that $\rho(n)\equiv \frac{1}{n}$, then 
\begin{eqnarray}
E_s^{ii}\equiv \frac{1}{n}\sum_{i=1}^{\sqrt{n}} i^2(1-\frac{1}{n})^{2i^2-2i+1}\equiv \frac{1}{n}\sum_{i=1}^{\sqrt{n}} i^2 \equiv \sqrt{n}.
\end{eqnarray}
Since $E_s^{ii}$ is increasing when  $\rho(n)$ is decreasing and its maximum possible order is $\sqrt{n}$, then $E_s^{ii}\equiv \sqrt{n}$ for all $\rho(n) \preceq \frac{1}{n}$.

For $\rho(n) \succ \frac{1}{n}$, we approximate the summation with the integral.

\footnotesize\begin{flalign}
&E_s^{ii}\equiv \frac{1}{n}\int_{v=1}^{\sqrt{n}} v^2(1-\rho(n))^{2v^2-2v+1}& \nonumber \\ 
&\equiv \{\frac{(1-\log(1-\rho(n)))\sqrt{2\pi(1-\rho(n))} erf(\frac{(2v-1)\sqrt{-\log(1-\rho(n))}}{\sqrt{2}})}{n\log^{3/2}(1-\rho(n))}& \nonumber \\
&+  \frac{-2\sqrt{-\log(1-\rho(n))}(2v+1)(1-\rho(n))^{2v^2-2v+1}}{n\log^{3/2}(1-\rho(n))}\}|_{v=1}^{\sqrt{n}}&
\end{flalign}\normalsize
where $erf$ is the error function which is always limited by $[-1,1]$ and is zero at zero.
If $\rho(n) \rightarrow 1$, then it is obvious that $E_s^{ii}\rightarrow 0$. For other values of $\rho(n) \succ \frac{1}{n}$ we use the third approximation in equation \eqref{eq:e_xn}, and also $-\log(1-\rho(n))\equiv \rho(n)$, which is true when $\rho(n)$ tends to zero while $n$ approaches infinity, and $-\log(1-\rho(n))\equiv 1$ for $\rho(n)\nrightarrow 0$ to prove that $E_s^{ii} \equiv \sqrt{n}$ for $\rho(n)\preceq \frac{1}{n}$, and $E_s^{ii} \equiv \frac{1}{n\rho^{3/2}(n)}$ for $\rho(n)\succ \frac{1}{n}$.
Since for $\rho(n)\preceq \frac{1}{n}$ the $E_s^{ii}$ reaches the maximum $E[h]$, therefore $E_c^{ii}$ cannot increase the scaling value of $E[h]$ anymore.
For $\rho \succ \frac{1}{n}$  we have $E_c^{ii} \equiv \sqrt{\frac{1}{ \rho(n)}}$.
Thus it can easily be verified that $E[h]\equiv E_s^{ii} \equiv \sqrt{n}$ for $\rho(n)\preceq \frac{1}{n}$, and $E[h]\equiv E_c^{ii}\equiv \sqrt{\frac{1}{ \rho(n)}}$ for $\rho(n)\succ \frac{1}{n}$.

Scenario $\mathbf{\romannumeral 3}$ - Let us define $E[h] = E_s^{iii}+E_c^{iii}$, where
\begin{eqnarray}\label{eq33}
&E_s^{iii}=& r^2(n)\sum_{i=2}^{\frac{1}{r(n)}} i^2(1-\rho(n))^{inr^2(n)} \nonumber \\ 
&E_c^{iii}=& r^2(n)(1-(1-\rho(n))^{nr^2(n)}) \nonumber \\
&&\times\{\sum_{i=2}^{\frac{1}{r(n)}}i\sum_{l=1}^{i-1}l(1-\rho(n))^{lnr^2(n)} \} 
	\end{eqnarray}
	
	First we check the behavior of $E_s^{iii}$ when $\rho(n)\equiv \frac{1}{nr(n)}$. Using the second region in equation \eqref{eq:e_xn} we will have $E_s^{iii}\equiv \frac{1}{r(n)}$.
	 $E_s^{iii}$ is increasing when $\rho(n)$ is decreasing and the maximum possible value for the number of hops is $\frac{1}{r(n)}$, then $E_s^{iii} \equiv \frac{1}{r(n)}$ for all $\rho(n) \preceq \frac{1}{nr(n)}$.
	
	By approximating the summation with integral, we arrive at
	\begin{eqnarray}
	&E_s^{iii}\equiv r^2(n)\int_2^{\frac{1}{r(n)}} v^2(1-\rho(n))^{vnr^2(n)}dv,& 
	\end{eqnarray}
	which equals to 
	
	\small\begin{eqnarray}\label{eq34}
	& \{(v^2\log^2 {(1-\rho(n))^{nr^2(n)}}-2v\log {(1-\rho(n))^{nr^2(n)}} +2)& \nonumber \\
	&\times\frac{r^2(n)(1-\rho(n))^{vnr^2(n)}}{\log^3 {(1-\rho(n))^{nr^2(n)}}}\}|_{v=2}^{\frac{1}{r(n)}}. &
	\end{eqnarray}\normalsize
	If $\frac{1}{nr(n)} \preceq \rho(n) \preceq \frac{1}{nr^2(n)}$, using the fact that $\log {(1-\rho(n))^{nr^2(n)}}\equiv -\rho(n)nr^2(n)$ and also equation \eqref{eq:e_xn}, we will have	$E_s^{iii}\equiv \frac{1}{n^3\rho^3(n) r^4(n)}.$
	
	When $\rho(n) \succeq \frac{1}{nr^2(n)}$, equation \eqref{eq34} tends to zero.

	Using the previous approximations along with $1-(1-\rho(n))^{nr^2(n)}\equiv 1$ for $\rho(n)\succeq \frac{1}{nr^2(n)}$, and $\rho(n)nr^2(n)$ for $\rho(n)\preceq \frac{1}{nr^2(n)}$, we can approximate $E_c^{iii}$ as its dominant terms ($E_c^{iii} \equiv \frac{1}{n\rho(n)}\sum_{i=2}^{\frac{1}{r(n)}} i \equiv \frac{1}{\rho(n)nr^2(n)}$).
	
	When $\rho(n)\succeq \frac{1}{nr^2(n)}$, the dominant term is $\Theta(1)$. Thus,
	
	\begin{equation}
	E[h]\equiv \left\{
		\begin{array}{ll}
					E_s^{iii} \equiv \frac{1}{r(n)}& \ \ \  \rho(n)\preceq \frac{1}{nr(n)} \\
					E_c^{iii}\equiv \frac{1}{\rho(n)nr^2(n)}& \ \ \  \frac{1}{nr(n)} \preceq \rho(n)\preceq \frac{1}{nr^2(n)} \\
					E_c^{iii}\equiv 1& \ \ \  \frac{1}{nr^2(n)} \preceq \rho(n)
			\end{array}\right .
	\end{equation}
	
	It can be seen that for large enough $\rho(n)$ the average number of hops between the nearest content location and the customer is just $\Theta(1)$ hops. This is the result of having $nr^2(n)$
	 caches in one hop distance for every requester. Each one of these caches can be  a potential source for the content. When the network grows, this number will increase and if $\rho(n)$ is large enough ($\frac{1}{nr^2(n)} \preceq \rho(n)$) the probability that at least one of these nodes contain the required data will approach 1, i.e., $\lim_{n\rightarrow \infty}  (1-(1-\rho(n))^{nr^2(n)}) = 1$.
	\end{IEEEproof}
 
\begin{IEEEproof}[Proof of Lemma \ref{lem:03}]
Assume that each content is retrieved with rate $\gamma$ bits/sec. The traffic generated because of one download from a cache (or server) at average distance of $E[h]$ hops from the requester node is $\gamma E[h]$. The total number of requests for a content in the network at any given time is limited by the number of nodes $n$. Thus the maximum total bandwidth needed to accomplish these downloads will be $nE[h]\gamma$, which is upper limited by ($\Theta(n)$) in scenarios $\mathbf{\romannumeral 1}$, $\mathbf{\romannumeral 2}$, and ($\Theta(\frac{1}{r^2(n)})$) in scenario $\mathbf{\romannumeral 3}$. Thus, $nE[h]\gamma \preceq  n$ and $\gamma_{max} \equiv \frac{1}{E[h]}$ in scenarios $\mathbf{\romannumeral 1}$, $\mathbf{\romannumeral 2}$, and $nE[h]\gamma \preceq  \frac{1}{r^2(n)}$ and $\gamma_{max} \equiv \frac{1}{E[h]nr^2(n)}$ 
in scenario $\mathbf{\romannumeral 3}$. Therefore the maximum download rate is easily derived using the results of Lemma \ref{lem:02}.
\end{IEEEproof}

\begin{IEEEproof}[Proof of Lemma \ref{lem:04}]
Each link between two nodes in scenarios $\mathbf{\romannumeral 1}$ and $\mathbf{\romannumeral 2}$, or two cells in scenario $\mathbf{\romannumeral 3}$ can carry at most $\Theta(1)$ bits per second. Here we calculate the maximum traffic passing through a link considering the throughput capacities derived in previous theorems, and check if any link can be a bottleneck.

Scenario $\mathbf{\romannumeral 1}$-  Each one of the four links connected to the server will carry all the traffic related to the items not found in the on-path caches. Thus, the total traffic related to item $k$ carried by each of those links is
$\psi_k=\sum_{i=1}^{\sqrt{n}}\gamma i (1-\rho^{(k)}(n))^i$.

When $\rho^{(k)}(n) \preceq \frac{1}{\sqrt{n}}$, we have $(1-\rho^{(k)}(n))^i \equiv 1$ for all $i\leq \sqrt{n}$. So this traffic is equal to $\psi_k=\sum_{i=1}^{\sqrt{n}}\gamma i \equiv n\gamma$.

When $\rho^{(k)}(n) \succeq \frac{1}{\sqrt{n}}$, using equation \eqref{eq:e_xn} the above summation can be written as
\begin{eqnarray}
&\gamma \frac{(-1+\rho^{(k)}(n))(\sqrt{n}\rho^{(k)}(n)(1-\rho^{(k)}(n))^{\sqrt{n}}+(1-\rho^{(k)}(n))^{\sqrt{n}}-1)}{(\rho^{(k)}(n))^2}& \nonumber \\
&\equiv  \frac{\gamma}{(\rho^{(k)}(n))^2}&. 
\end{eqnarray}

The total traffic is $\psi=\sum_{k=1}^m \alpha_k \psi_k$ which must be less than one. If $\rho^{(k)}(n) \succeq \frac{1}{\sqrt{n}}$ for all the items, then the item with minimum $\rho^{(k)}(n)$ will be the dominant factor in the above equation ($\psi \equiv \Theta(\frac{\gamma}{\underset{k}{min}(\rho^{(k)}(n))^2)})$), and if at least one item has $\rho^{(k)}(n) \preceq \frac{1}{\sqrt{n}}$, it will put the bound on the maximum rate ($\psi \equiv n\gamma$).
Thus, $\psi \equiv min(n\gamma,\frac{\gamma}{\underset{k}{min}(\rho^{(k)}(n))^2)}) \preceq 1$, then $\gamma_{max}  \equiv max(\frac{1}{n}, \underset{k}{min}((\rho^{(k)}(n))^2))$.

Therefore, the links directly connected to the server will be a bottleneck if $\gamma$ is more than the above values. On the other hand, the traffic related to item $k$ carried by a node to cache content in level $j$ is $\sum_{i=1}^{\sqrt{n}-j}\gamma i(1-\rho^{(k)}(n))^i \preceq \sum_{i=1}^{\sqrt{n}}\gamma i(1-\rho^{(k)}(n))^i$, so the server links carry the maximum load, and thus the derived upper limits are supportable in every link. 

Scenario $\mathbf{\romannumeral 2}$- Each one of the four links connected to the server will carry all the traffic related to the items not found in any caches closer to the requester. Thus, the total traffic related to item $k$ ($\psi_k$) carried by each of those links is
\begin{eqnarray}
&\gamma (1-\rho^{(k)}(n))+\sum_{i=1}^{\sqrt{n}}4\gamma i (1-\rho^{(k)}(n))^{(1+4\sum_{j=1}^ij)}& \nonumber \\
&\equiv \gamma (1-\rho^{(k)}(n))+\sum_{i=1}^{\sqrt{n}}\gamma i (1-\rho^{(k)}(n))^{2i^2+2i+1},& \nonumber \\
&\equiv \gamma\{(1-\rho^{(k)}(n)) +
 \frac{(1-\rho^{(k)}(n))^n-(1-\rho^{(k)}(n))^4 }{\log (1-\rho^{(k)}(n))/(1-\rho^{(k)}(n))}& \nonumber \\
&+\frac{\sqrt{-\frac{\log (1-\rho^{(k)}(n))}{1-\rho^{(k)}(n)}}erf(\sqrt{-n\log (1-\rho^{(k)}(n))})}{\log (1-\rho^{(k)}(n))/(1-\rho^{(k)}(n))}& \nonumber \\
&-\frac{\sqrt{-\frac{\log (1-\rho^{(k)}(n))}{1-\rho^{(k)}(n)}}erf(\sqrt{-\log (1-\rho^{(k)}(n))})}{\log (1-\rho^{(k)}(n))/(1-\rho^{(k)}(n))}\}. & 
\end{eqnarray} 

If $\rho^{(k)}(n) \preceq \frac{1}{n}$, then $(1-\rho^{(k)}(n))^{2i^2+2i+1} \equiv 1$ for all $1\leq i \leq \sqrt{n}$. Thus the above traffic will be $\psi_k \equiv n\gamma$. If $\rho^{(k)}(n) \succeq \frac{1}{n}$ the above equation is equivalent to $\psi_k \equiv \frac{\gamma}{\rho^{(k)}(n)}$.

The total traffic then is $\psi \equiv \sum_{k=1}^m \alpha_k \psi_k \preceq 1$. If $\rho^{(k)}(n) \succeq \frac{1}{n}$ for all the items, then $\psi \equiv \frac{\gamma}{\underset{k}{min}(\rho^{(k)}(n))}$. If $\rho^{(k)}(n) \preceq \frac{1}{n}$ for at least one item, then $\psi \equiv n\gamma$.  Thus, $\psi \equiv min(n\gamma,\frac{\gamma}{\underset{k}{min}(\rho^{(k)}(n))}) \preceq 1$, then $\gamma_{max}  \equiv max(\frac{1}{n}, \underset{k}{min}(\rho^{(k)}(n)))$.
 
 Using similar reasoning as in scenario $\mathbf{\romannumeral 2}$ other links carry less traffic, so the above capacities are supportable for all the other links.
 
Scenario $\mathbf{\romannumeral 3}$- The traffic load for item $k$ between the server cell and each of the four neighbor cells ($\psi_k$) is given by
\begin{eqnarray}
&&\gamma nr^2(n) \{(1-\rho^{(k)}(n))+\sum_{i=2}^{\frac{1}{r(n)}}i(1-\rho^{(k)}(n))^{inr^2(n)}\} \nonumber \\
&\equiv& \gamma nr^2(n) \{(1-\rho^{(k)}(n)) \nonumber \\
&+&\frac{(1-\rho^{(k)}(n))^{nr(n)}(nr(n)\log (1-\rho^{(k)}(n))-1)}{\log^2 (1-\rho^{(k)}(n))^{nr^2(n)}} \nonumber \\
&-&\frac{(1-\rho^{(k)}(n))^{nr^2(n)}(\log (1-\rho^{(k)}(n))^{nr^2(n)}-1)}{\log^2 (1-\rho^{(k)}(n))^{nr^2(n)}}\} \label{eq44}
\end{eqnarray}

If $\rho^{(k)}(n) \preceq \frac{1}{nr(n)}$, then $(1-\rho^{(k)}(n))^{inr^2(n)}\rightarrow 1$ for $2\leq i \leq \frac{1}{r(n)}$, thus the traffic load equals to $\gamma nr^2(n) \sum_{i=2}^{\frac{1}{r(n)}}i \equiv n\gamma $. 

If $\frac{1}{nr(n)} \preceq \rho^{(k)}(n) \preceq \frac{1}{nr^2(n)}$, then the maximum traffic load $\psi_k$ on a link is 
\begin{eqnarray}
 &&\gamma nr^2(n) + \gamma nr^2(n) \frac{1 + 2 \rho^{(k)}(n) nr^2(n)}{(\rho^{(k)}(n))^2 n^2r^4(n)} \nonumber \\
&\equiv& \frac{\gamma}{(\rho^{(k)}(n))^2 nr^2(n)}
\end{eqnarray}

If $\rho^{(k)}(n) \succeq \frac{1}{nr^2(n)}$, then equation \eqref{eq44}  is equivalent to $\gamma nr^2(n)$. Therefore, if $\rho^{(k)}(n) \succeq \frac{1}{nr^2(n)}$ for all the items, then the total traffic ($\psi = \sum_{k=1}^m \alpha_k \psi_k$) is simply $\psi \equiv \gamma nr^2(n)$. If $\rho^{(k)}(n) \preceq \frac{1}{nr(n)}$ for all items but there is at least one item for which $\rho^{(k)}(n) \preceq \frac{1}{nr^2(n)}$, then the total traffic is dominated by the traffic generated by the item with the least $\rho^{(k)}(n)$ ($\rho^{(k)}(n) \preceq \frac{1}{nr^2(n)}$). And finally if there is at least one item for which $\rho^{(k)}(n) \preceq \frac{1}{nr(n)}$, then it will generate the dominant traffic ($\psi \equiv n\gamma$). Thus, $\psi \equiv min[n\gamma,$  $max(\gamma nr^2(n), \frac{\gamma}{\underset{k}{min}(\rho^{(k)}(n))^2 nr^2(n)})] \preceq 1$, $\gamma_{max} \preceq max[\frac{1}{n},min(\frac{1}{nr^2(n)},\underset{k}{min}((\rho^{(k)}(n))^2) nr^2(n))].$ 
Note that if there is no cache in the system, or $\rho(n)$ is very low, less than the stated threshold values, almost all the requests would be served by the server, and the maximum download rate would be $\Theta(\frac{1}{n})$.

\end{IEEEproof}

\bibliographystyle{IEEEtran}
\vspace{-0.6em}
\bibliography{IEEEabrv,TCOM-TPS-16-0233_R1}

\end{document}